%% file: main.tex
\documentclass[10pt,leqno]{amsart}
\usepackage[T1]{fontenc}
\usepackage{lmodern}
\usepackage{graphicx}
\usepackage{indentfirst,csquotes}

\usepackage{amssymb,amsthm,amsmath}
\usepackage{thmtools,thm-restate}
\usepackage[dvipsnames]{xcolor}
\usepackage{comment,listings,paralist,etoolbox}
\usepackage{hyperref}
\newtheorem{theorem}{Theorem}
\newtheorem{definition}[theorem]{Definition}

\newtheorem{lemma}[theorem]{Lemma}

\newtheorem{corollary}[theorem]{Corollary}

\hypersetup{ colorlinks=true, linkcolor=black, filecolor=black, urlcolor=black }

\title{Collaborative Package Delivery in the Plane} 

\author[K. C. Huynh]{Kien C. Huynh}
\address{Communications and Transport Systems, ITN, Link\"oping University, Sweden}
\email{kchuynh@cs.stonybrook.edu}

\author[J. S. B. Mitchell]{Joseph S. B. Mitchell}
\address{Department of Applied Mathematics and Statistics, Stony Brook University, USA}
\email{joseph.mitchell@stonybrook.edu}

\date{\today}

\begin{document}

\begin{abstract}
    In this paper, we study the package delivery problem with $n$ heterogeneous collaborative drones in 2D. Given $n$ drones in the plane, each with a different parameter, the task is to find a schedule to deliver a package from $s$ to $t$ to satisfy certain objectives, and a drone is allowed to hand off the package to another drone.
    We investigate two natural versions of the problem: fast delivery (\textsc{FD}) -- drones could have different speeds, and we want to minimize total delivery time; and energy-efficient delivery (\textsc{ED}) -- drones have different energy consumption rates, and we want to minimize total energy spent.
    For the \textsc{FD} version, we show that the problem is algebraically hard in 2D even with just two drones, i.e., solving some instances equates to finding the roots of polynomials that have no solutions in radicals.
    We prove that a strategy in which the package is optimally moved along the $s-t$ line segment is a 2-approximation to general movement, and provide an FPTAS without movement restriction.
    We also show that the FPTAS applies almost immediately to \textsc{ED} after a few tweaks.
\end{abstract}

\maketitle

\section{Introduction}

The advancement of drone technology has led to many exciting applications in recent years.
In particular, drones have been used for many different tasks in a wide variety of areas such as logistics, search and rescue, and entertainment.
A promising usage of drones in logistics is for quick deliveries, which has been receiving a lot of development in both theoretical and practical aspects.
One specific variant of delivery with drones, often called package delivery with a fleet of collaborative drones, involves coordinating multiple drones to complete a delivery while minimizing time (makespan), total energy spent, or some other objectives.
The interesting aspect of this problem is that the drone can hand the package to another for a more efficient schedule, which makes the solution non-trivial. 
The problem type was first posed as an open question in \cite{anaya2012collecting}, under the theme of message passing between robots, and since then has gained considerable interest in the algorithmic community with many different variants.
Most work on this delivery problem was for the graph domain (\cite{bartschi2017energy,bartschi2018collective,carvalho2021fast}), while results have been lacking for the continuum.
Specifically, the Euclidean variant was raised as an open problem in \cite{carvalho2021fast}.
So far, we have only found one related study that has attempted to solve this problem in the plane \cite{coleman2021message}, for both the offline and online versions of fast delivery.
We address this gap by providing new results on the problem of delivery in the plane, which, in contrast to the graph version, poses a unique set of challenges.

\textbf{Related work.}
\input{related_work}



\textbf{Our contributions.} We present several results for delivery with collaborative drones in the plane.
In particular:
\begin{itemize}
    \item We give a $\Theta(n\log n)$ 2-approximation, and the first FPTAS for fast delivery (\textsc{FD}).
    We remark that the FPTAS can be modified for 2D domains with polygonal obstacles.
    The only prior work in the plane \cite{coleman2021message} gives a $(\sqrt{2}+\epsilon)$-approximation with a running time depending on $\Delta$, a parameter derived from the diameter of the input points.
    \item We prove that FD in the plane is algebraically hard with just two drones. In contrast, the optimum in the graph version can be determined by a rational function of the input (even if the exchange point can be in the interior of some edge) \cite{carvalho2021fast}. 
    \footnote{Although not stated explicitly in~\cite{carvalho2021fast}, this follows from their construction: agent movements are represented as lines $y = t_i \pm x/v_i$ in a time-distance diagram (here, $t_i$ is the arrival time of drone $i$ at some vertex of the graph), and every exchange point coordinate is an intersection of two such lines, hence a rational expression in the speeds and arrival times. Since arrival times are quotients of edge-length sums by speeds, an induction over Dijkstra’s recursion shows that every exchange point coordinate and the optimal makespan lie in $\mathbb{Q}(\{l_e\},\{v_i\})$ ($l_e$ is the weight of edge $e$); no algebraic extension of the input field arises.}
    \item The same FPTAS design can also be applied to energy-efficient delivery (\textsc{ED}). To our knowledge, this is also the first FPTAS for \textsc{ED} in the plane.
\end{itemize}

\textbf{Preliminaries.} We are given a set of $n$ drones, a package located at $s$ to be delivered to $t$, where $s, t\in \mathbb R^2$.
Each drone $i$, $1 \le i \le n$, has a different starting location $p_i \in \mathbb{R}^2$ and a parameter $v_i$ (speed) or $r_i$ (energy consumption rate), depending on which version we are working with.
W.l.o.g., we assume that $s=(0,0)$, $t = (x_t, 0), x_t > 0,$ and $v_1 < v_2 < \ldots < v_n$ (and similarly, for $\{r_i\}_{i=1}^n$).
The package can be picked up at $s$ by any drone, and can be handed over from one drone to another.
Since we only have one package, we can assume that the package will only be handed from a less efficient drone (slower or burns more energy) to a more efficient drone.
We assume that handoffs occur instantly and cost zero time or energy.

A solution is a pair of sequences $S=(\Pi, Q)$: $\Pi=(\pi_1,\pi_2,\ldots \pi_m)$ where $\pi_j$ is a drone index, and $Q = (q_0=s,q_1,q_2,\ldots q_{m}=t)$ is the sequence of exchange points in $\mathbb R^2$.
In any solution, the drone $\pi_j$ picks up the package from $q_{j-1}$ and gives it to drone $\pi_{j+1}$ at $q_j$.
Let $d(\cdot, \cdot)$ be the Euclidean distance between two points in $\mathbb R^2$. 
For \textsc{FD}, the time that it takes for the first drone in $\Pi$ to pick up the package at $s$ and move to $q_1$ is
\[
\tau_S(1) = \frac{1}{v_{\pi_1}}\big(d(p_{\pi_1}, s) + d(s, q_{1})\big),
\]
and the time for drone $\pi_j$ to take the package from $q_{j-1}$ to $q_j$ is
\[
\tau_S(j) = \max \bigg( \frac{1}{v_{\pi_j}}d(p_{\pi_j}, q_{j-1}), \tau_S(j-1)  \bigg) + \frac{1}{v_{\pi_j}}d(q_{j-1}, q_{j}).
\]
$\tau_S(m)$ is the total makespan to deliver the package to $t$.
The function $\max$ in the above recursion essentially means that the one between drones $\pi_{j-1}$ and $\pi_j$ that reaches $q_{j-1}$ first will have to wait for the other.
The problem \textsc{FD} asks for a delivery schedule $S$ with the minimum $\tau_S(m)$ in the plane.
We show the respective objective function for \textsc{ED} in Section~\ref{sect:ED}.

\section{ Fast delivery with collaborative drones}\label{sect:FD}



First, we will define canonical optimal solutions for \textsc{FD}:
\begin{definition}\label{def:canonical}
    A solution $S = (\Pi, Q)$ is \emph{canonical} if
    (i) every drone $\pi_j \in \Pi$ departs at time $0$ and moves without
    stopping until it goes to $q_{j-1}$, and
    (ii) for every $j > 1$, drones $\pi_{j-1}$ and $\pi_j$ arrive at
    $q_{j-1}$ at the same time.
\end{definition}
We claim that there always exists a canonical optimal solution (see proof in Appendix~\ref{app:canonical}):
\begin{restatable}{lemma}{canonical}
\label{lem:canonical}
There exists a canonical optimal solution for \textsc{FD}.
\end{restatable}
We will assume that all optimal schedules mentioned from here on are canonical, and denote them by $S^*$. 
We want to argue for the existence of $S^*$ because in $S^*$, each drone $\pi_j$ arrives at $q_{j-1}$ at the same time the package does, i.e., $\frac{1}{v_{\pi_j}}d(p_{\pi_j}, q_{j-1}) = \tau_{S^*}(j-1)$. This means the $\max$ in the recursion can be dropped:
$$\tau_{S^*}(j) = \frac{1}{v_{\pi_j}}d(p_{\pi_j}, q_{j-1}) + \frac{1}{v_{\pi_j}}d(q_{j-1}, q_{j}).$$
For all $j >1$:

\begin{lemma}\label{lem:never-too-far}
    From time $0$ until $\tau_{S^*}(j-1)$, the package never moves more than $d(p_{\pi_j}, q_{j-1})$ from $s$.
\end{lemma}
Since $S^*$ is canonical, $\tau_{S^*}(j-1) = \frac{1}{v_{\pi_j}}d(p_{\pi_j}, q_{j-1})$, so the package has traveled for a duration of $\frac{1}{v_{\pi_j}}d(p_{\pi_j}, q_{j-1})$. During this time, no drone $\pi_{j'}$ with $j' < j$ can move the package farther than $v_{\pi_j} \cdot \frac{1}{v_{\pi_j}}d(p_{\pi_j}, q_{j-1}) = d(p_{\pi_j}, q_{j-1})$ from $s$, since $v_{\pi_{j'}} < v_{\pi_j}$.



\subsection{2-approximation algorithm}\label{subsec:2-approx}

The 2-approximation works by forcing all handoffs to occur in the line segment connecting $s$ and $t$, which can be computed in $\Theta(n \log n)$ \cite{blank2025algorithms}.

Consider the sequence of optimal exchange points $Q^*$ and their projections on the $x$-axis, $Q^x =\{q^x_j = (x_{q_j},0) | q_j \in Q^*\}$.
The solution $S^x = (\Pi^*, Q^x)$ is a sequence of handoffs where the drone $\pi_j$ goes to the projected point $q^x_{j-1}$ to pick up the package and move it to $q^x_j$.
Naturally, some waiting may occur at these new exchange points (unless $Q^x = Q^*$).
We will prove the following upper bound of the makespan of $S^x$:

\begin{lemma}\label{lem:2tau2}
    $\tau_{S^x}(m) \le 2\tau_{S^*}(m) $.
\end{lemma}

\begin{proof}
The makespan recursion using $S^x$ at the $j$-th handoff is:
\[
\tau_{S^x}(j) = \max \bigg( \frac{1}{v_{\pi_j}}d(p_{\pi_j}, q^x_{j-1}), \tau_{S^x}(j-1)  \bigg) + \frac{1}{v_{\pi_j}}d(q^x_{j-1}, q^x_{j}).
\]
First, notice that $d(p_{\pi_j}, q^x_{j-1}) \le d(p_{\pi_j}, q_{j-1}) + d(q_{j-1}, q^x_{j-1})$ due to triangle inequality (note that $d(q_{j-1}, q^x_{j-1})$ is the distance from $q_{j-1}$ to the $x$-axis).
By Lemma~\ref{lem:never-too-far} (applied to the optimal canonical solution $S^*$), $d(s, q_{j-1}) \le d(p_{\pi_j}, q_{j-1})$. Since $q^x_{j-1}$ is the projection of $q_{j-1}$ onto the $x$-axis and $s$ lies on the $x$-axis, we have $d(q_{j-1}, q^x_{j-1}) \le d(q_{j-1}, s) \le d(p_{\pi_j}, q_{j-1})$, so $d(p_{\pi_j}, q^x_{j-1}) \le 2d(p_{\pi_j}, q_{j-1})$.
Now, consider how the package travels from $s$ to $t$: the package moves from one exchange point ($q^x_{j-1}$) to another ($q^x_{j}$) until it reaches $t$.
Each time it reaches a projected exchange point $q^x_{j-1}$, the package must either (1) wait there until $\pi_j$ picks it up (since $\pi_{j-1}$ reached $q^x_{j-1}$ before $\pi_j$), or (2) be instantly carried by $\pi_j$ if $\pi_j$ arrived before or at the same time as $\pi_{j-1}$.
Among all projected exchange points, let $q^x_{k-1}$ be the last exchange point such that case (1) occurs.
Then, the makespan of $S^x$ can be written as
\[
\tau_{S^x}(m) = \frac{1}{v_{\pi_k}}d(p_{\pi_k}, q^x_{k-1}) + \sum_{j=k}^{m}\frac{1}{v_{\pi_j}}d(q^x_{j-1}, q^x_{j}).
\]
Intuitively, this is because the $\max$ function in the recursion allows the ``lateness'' of drone $k$ to overcome the lateness of all previous drones: only the delay caused by the last late drone, $k$, matters.
Since $d(p_{\pi_k}, q^x_{k-1}) \le 2d(p_{\pi_k}, q_{k-1})$ and $d(q^x_{j-1}, q^x_{j}) \le d(q_{j-1}, q_{j})$, $\tau_{S^x}(m) $ is at most
\[
\frac{2}{v_{\pi_k}}d(p_{\pi_k}, q_{k-1}) + \sum_{j=k}^{m}\frac{1}{v_{\pi_j}}d(q_{j-1}, q_{j}) = \frac{1}{v_{\pi_k}}d(p_{\pi_k}, q_{k-1}) + \tau_{S^*}(m) \le 2\tau_{S^*}(m).
\]
The equality is true because, based on Lemma~\ref{lem:canonical}, we can write:

\begin{align*}
    \tau_{S^*}(m) &= \frac{1}{v_{\pi_1}}d(p_{\pi_1}, q_{0}) + \sum_{j=1}^m \frac{1}{v_{\pi_j}}d(q_{j-1}, q_{j}) \\
    &= \Big(\frac{1}{v_{\pi_1}}d(p_{\pi_1}, q_{0}) + \sum_{j=1}^{k-1} \frac{1}{v_{\pi_j}}d(q_{j-1}, q_{j})\Big) + \sum_{j=k}^m \frac{1}{v_{\pi_j}}d(q_{j-1}, q_{j}) \\
    &= \tau_{S^*}(k-1) + \sum_{j=k}^m \frac{1}{v_{\pi_j}}d(q_{j-1}, q_{j}) = \frac{1}{v_{\pi_k}}d(p_{\pi_k}, q_{k-1})+ \sum_{j=k}^m \frac{1}{v_{\pi_j}}d(q_{j-1}, q_{j}).
\end{align*}
\end{proof}
Consider the drone delivery problem with the restriction that drones can only perform handoffs on the $x$-axis.
We will use $S^ {(2)}=(\Pi^ {(2)}, Q^ {(2)})$ to denote the optimal solution of this restricted version, and $\tau^{(2)}$ to denote its makespan.
It follows that $\tau^{(2)} \le \tau_{S^x}(m)$.
To find $S^ {(2)}$, we claim that (the proof is shown in Appendix~\ref{app:claim1}):
\begin{restatable}{claim}{neverbackward}
    The package does not travel toward the negative direction of the $x$-axis in $S^ {(2)}$.
\end{restatable}
With this, we can assume that the package only needs to travel between $s$ and $t$, and we can then find $S^ {(2)}=(\Pi^ {(2)}, Q^ {(2)})$ in $\Theta(n \log n)$ using the algorithm in Section 4.2 of \cite{blank2025algorithms}.

\begin{theorem}
    There exists a $\Theta(n \log n)$ algorithm that returns a 2-approximation for \textsc{FD}.
\end{theorem}

\subsection{FPTAS}
A naive approach to designing the FPTAS is to create a uniform grid on top of the bounding box (or convex hull) of the given positions, and set the cell side length to be some $\epsilon > 0$.
The problem with this is that it only gives a pseudo-polynomial algorithm since the running time (which depends on the number of grid points) will scale with $\big(\frac{D}{\epsilon}\big)^2$ where $D$ is the largest distance between two points in the input.
$D$ can be arbitrarily longer than the trajectory of the package in $OPT$, e.g., when $D$ belongs to the starting positions of drones irrelevant to $OPT$.
Instead, we can obtain a boundary for the exchange points that actually depends on $O(OPT)$ using Lemma~\ref{lem:never-too-far} and Lemma~\ref{lem:2tau2}.
Another quirk of the problem is that there cannot be one grid with a fixed cell side length that can be used for all drones. 
Suppose the universal cell side length is $O(\epsilon)$; snapping either exchange point of $v_j$ to a grid vertex will introduce an error bound that scales with $\frac{\epsilon}{v_j}$: large $\epsilon$ will result in large errors for slow drones, but setting $\epsilon$ too small (e.g., to accommodate $v_1$) will be wasteful for fast drones, and the grid size will scale with $\big(\frac{v_n}{v_1}\big)^2$, which can be arbitrarily large.
We avoid this issue by designing a speed-dependent family of grids in which the total number of points polynomially depends on $n$, $\log\big(\frac{v_n}{v_1}\big)$, and $\epsilon$.

We first explain our idea of creating one grid point set for each drone in the input ($n$ grids in total), and will later explain how we can reduce the number of grids when there are drones with similar speeds in the input.
We define $G_i$ to be the axis-parallel square grid centered at $s$ for drone $i$, $V_i$ its vertex set, and $L_i$ the side length of the grid.
The set of vertices $V=\big(\bigcup_{1\le i \le n} V_i\big) \cup \{s,t,p_1,\ldots p_n\}$ will serve as candidate points for package exchange.
Since this candidate set is discrete, we can apply known algorithms on graphs \cite{carvalho2021fast} to find the optimal solution on the complete graph over $V$.
For the optimal solution on $V$ to be an $(1+\epsilon)$-approximation, we need to (1) bound the side lengths and (2) make each grid fine-grained enough.
We first describe how we bound the side lengths $L_i$ for each grid so that the bounding box of $G_i$ surely contains both exchange points of drone $i$ (if drone $i$ participates in $S^*$).
This can be achieved by setting $L_i = 2 \tau^{(2)}v_i$.
\begin{lemma}
    If $i\in \Pi^*$, then both of its exchange points are in the bounding box of $V_i$.
\end{lemma}
\begin{proof}
    Assume $i = \pi_j \in \Pi^*$.
    Recall Lemma~\ref{lem:never-too-far}, which implies that after drone $\pi_j$'s participation (pickup at $q_{j-1}$ and handoff at $q_j$), the package could never travel further than $d(p_{\pi_j}, q_{j-1}) + d(q_{j-1}, q_j)$ from $s$ at time $\tau_{S^*}(j)$.
    Since $d(p_{\pi_j}, q_{j-1}) + d(q_{j-1}, q_j) = \tau_{S^*}(j)v_{\pi_j} \le \tau_{S^*}(m)v_{\pi_j} \le \tau^{(2)}v_{\pi_j}$, by setting the grid length $L_{\pi_j} = 2 \tau^{(2)}v_{\pi_j}$, we can guarantee that both $q_{j-1}$ and $q_{j}$ are contained in the bounding box of $V_{\pi_j}$.
\end{proof}

The next step is to make $G_i$ fine-grained enough.
Formally, we define $G_i$ as the grid obtained by splitting a square of side length $L_i$ into $M \times M$ cells.
Then $(M + 1)^2 = |V_i|$ and the side length of each cell is $\frac{L_i}{M} = \delta_i$.
Suppose that $S^V = \big(\Pi ^V, Q^V\big )$ is the optimal solution on $V$, and $|\Pi ^V| = m^V$.
If we set $M= \Big\lceil \frac{(n+1)6\sqrt{2}}{\epsilon}\Big\rceil$, we can prove the following:
\begin{lemma}\label{lemma:snapped-fd}
$\tau_{S^V}(m^V) \le (1 + \epsilon) \tau_{S^*}(m)$.
\end{lemma}
\begin{proof}
Consider the optimal solution $S^*=(\Pi^*, Q^*)$ in $\mathbb {R} ^2$.
We will argue that we can take a subset of drones $\Pi' \subseteq \Pi^*$, and use exchange points in $V$ to produce a good approximation for $S^*$.
$\Pi'$ is defined as follows:
\begin{itemize}
    \item For drone $\pi_1 \in \Pi^*$, its first exchange point $q_0 (=s)$ will be contained within a cell $\gamma_1$ in the grid $G_{\pi_1}$.
    Consider the maximal sequence of exchange points $(q_0,q_1, q_2, \ldots,q_{j-1})$ that belong to $\gamma_1$, i.e., $q_{j} \notin \gamma_1$.
    We put $\pi_j$ in $\Pi'$ while ignoring the rest.
    Notice that drone $\pi_j$ is the drone that carries the package out of cell $\gamma_1$ to some exchange point $q_{j}$, and $\pi_j$ is also the fastest drone among all drones involved in the package exchange sequence in $\gamma_1$. In some cases, $\pi_j$ can be $\pi_1$.
    \item Start from $\pi_j$, and execute the same process with the cell $\gamma_j$ in $G_{\pi_j}$ containing $q_{j}$: we ignore all drones both of whose exchange points are contained in $\gamma_j$ and only keep the drones that enter or exit the cell.
    \item Repeat the above step until no drone is left. Note that in the last cell, we will always keep $\pi_m$ and use it to deliver to $t$ (regardless of whether $t$ is in the cell or not).
\end{itemize}
Next, we will show how we snap the exchange points of the remaining drones to the vertices of $V$.
Consider a cell $\gamma$ where some package exchanges occur inside; let $\pi_\text{in}$ (resp. $\pi_\text{out}$) be the drone entering (resp. exiting) the cell (there could be many cells where $\pi_\text{in}$ enters and $\pi_\text{out}$ exits; but $\gamma$ is one with the smallest size based on the pruning process).
Both $\pi_\text{in}$ and $\pi_\text{out}$ belong to $\Pi'$ after the process described above.
Let $q_\text{in}$ be the location where $\pi_\text{in}$ hands over the package to some faster drone, and $q_\text{out}$ be the location where $\pi_\text{out}$ picks up the package from some slower drones.
Excluding $s$ and $t$ (which will stay the same), we snap both $q_\text{in}$ and $q_\text{out}$ to the grid vertex $q' \in V_{\pi_\text{in}}$ closest to $q_\text{in}$ and use $q'$ as the exchange point between $\pi_\text{in}$ and $\pi_\text{out}$.
We will use these snapped vertices in our new sequence of exchange points $Q' =(s=q'_0 ,q'_1, \ldots$ $ q'_{\mu}=t)$.
Let $\Pi' = (\pi'_1,\ldots \pi'_{\mu})$, we denote the solution $(\Pi',Q')$ as $S'$. 
We will denote $a_j$ and $b_j$ as the original pickup and handoff points of $\pi'_j$ before snapping them to $V$.
Note that $a_j$ is not necessarily equal to $b_{j-1}$ after pruning.
Using a similar argument in the proof of Lemma~\ref{lem:2tau2}, the makespan of $S'$ depends on the ``last'' drone $\pi'_k \in \Pi'$ such that the package arrives at $q'_{k-1}$ before $\pi'_k$ does, which can be written as
\[
\tau_{S'}(\mu) = \frac{1}{v_{\pi'_k}}d(p_{\pi'_k}, q'_{k-1}) + \sum_{j=k}^{\mu}\frac{1}{v_{\pi'_j}}d(q'_{j-1}, q'_{j}).
\]

First, observe that $q'_{k-1}$ belongs to a cell of the grid $G_{\pi'_{k-1}}$.
Since the side length of such a cell is $\delta_{\pi'_{k-1}}$, the duration of the new trajectory of drone $\pi'_k$ is longer than that of the original path in $S^*$ by an additive factor of at most
\[
\frac{\delta_{\pi'_{k-1}}\sqrt{2}}{v_{\pi'_k}} \le \frac{\delta_{\pi'_{k}}\sqrt{2}}{v_{\pi'_{k}}} = \frac{L_{\pi'_{k}} \sqrt{2}}{Mv_{\pi'_{k}}} = \frac{\tau^{(2)}2\sqrt{2}}{M}.
\]
In case $k=1$, we define $G_{\pi'_0} = G_{\pi'_1}$. Then:
\begin{equation}\label{eq:1}
    \frac{1}{v_{\pi'_k}}d(p_{\pi'_k}, q'_{k-1}) \le \frac{1}{v_{\pi'_k}}d(p_{\pi'_k}, a_k) +\frac{\tau^{(2)}2\sqrt{2}}{M}.
\end{equation}

Similarly, $\forall j\ge k$:
\begin{align*}
\frac{d(q'_{j-1}, q'_{j})}{v_{\pi'_j}} \le \frac{\delta_{\pi'_{j-1}}\sqrt{2}}{v_{\pi'_j}} + \frac{d(a_j, b_j)}{v_{\pi'_j}} + \frac{\delta_{\pi'_{j}}\sqrt{2}}{2v_{\pi'_j}}\le \frac{d(a_j, b_j)}{v_{\pi'_j}} + \frac{\tau^{(2)}3\sqrt{2}}{M}.
\end{align*}

(Snapping $a_j$ to $q'_{j-1}$ costs $\frac{\delta_{\pi'_{j-1}}\sqrt{2}}{v_{\pi'_j}} $ but snapping $b_j$ to $q'_j$ costs $\frac{\delta_{\pi'_{j}}\sqrt{2}}{2v_{\pi'_j}} $; recall that we snap $q_{in}$ (i.e., $b_j$) to the closest grid vertex.)
We now bound the second term of $\tau_{S'}(\mu) $:
\begin{equation}\label{eq:2}
\sum_{j=k}^{\mu}\frac{1}{v_{\pi'_j}}d(q'_{j-1}, q'_{j}) \le \sum_{j=k}^{\mu}\Bigg(\frac{1}{v_{\pi'_j}}d(a_j, b_j) + \frac{\tau^{(2)}3\sqrt{2}}{M}\Bigg).
\end{equation}

By summing (\ref{eq:1}) and (\ref{eq:2}):
\begin{align*}
    \tau_{S'}(\mu) &\le  \frac{1}{v_{\pi'_k}}d(p_{\pi'_k}, a_k) + \sum_{j=k}^{\mu}\Bigg(\frac{1}{v_{\pi'_j}}d(a_j, b_j)\Bigg) +\frac{\tau^{(2)}2\sqrt{2}}{M} + \sum_{j=k}^{\mu} \frac{\tau^{(2)}3\sqrt{2}}{M}\\
    &\le  \tau_{S^*}(m) + (\mu-k+2)\frac{\tau^{(2)}3\sqrt{2}}{M} \le \tau_{S^*}(m)\Bigg(1 + \frac{(m+1)6\sqrt{2}}{M} \Bigg).
\end{align*}

Note that $\frac{1}{v_{\pi'_k}}d(p_{\pi'_k}, a_k) + \sum_{j=k}^{\mu}\Big(\frac{1}{v_{\pi'_j}}d(a_j, b_j)\Big) \le \tau_{S^*}(m)$ because the left-hand side is the sum of the subset of terms taken from $\tau_{S^*}(m)$ that survive after drone-pruning.
By setting $M = \Big\lceil \frac{(n+1)6\sqrt{2}}{\epsilon} \Big\rceil$ (since $m$ is actually unknown), we can achieve an $(1+\epsilon)$-approximation with $S'$, as well as $S^V$, since it is a solution at least as good as $S'$.
\end{proof}

We will now explain how we reduce the number of grids by grouping similar speeds into buckets.
Let $\alpha > 0$ be a parameter; we will create a set of intervals of speeds $I_v=\{[v_1(1+\alpha)^\ell, v_1(1+\alpha)^{\ell+1}]| 0 \le \ell < \beta \}$, where $\beta = \big \lceil \log_{1+\alpha}\frac{v_n}{v_1} \big \rceil = O\big(\frac{1}{\alpha}\log \frac{v_n}{v_1} \big )$.
An interval is non-empty if it contains at least the speed of one drone in the input.
For each non-empty interval, we create a grid with side length $L_\ell = 2\tau^{(2)}v_1(1+\alpha)^{\ell+1}$, which is guaranteed to contain all exchange points of any drone in the $\ell$-th bucket.
Let $\delta_\ell$ be the length of a cell where an exchange occurs for drone $\pi'_j$; snapping the optimal exchange point to a vertex of the cell will introduce a delay of at most
$\frac{\delta_{\ell}\sqrt{2}}{v_{\pi'_j}}  = \frac{L_{\ell} \sqrt{2}}{Mv_{\pi'_{j}}} \le \frac{\tau^{(2)}(1+\alpha)2\sqrt{2}}{M}$
(The inequality uses $v_1(1+\alpha)^\ell \le v_{\pi'_j} \le v_1(1+\alpha)^{\ell+1}$ for some $\ell$).
The final makespan is:
\begin{align*}
    \tau_{S'}(\mu) &\le  \tau_{S^*}(m)\Bigg(1 + \frac{(m+1)(1+\alpha)6\sqrt{2}}{M} \Bigg) \le \tau_{S^*}(m)(1 + \epsilon(1+\alpha) ).
\end{align*}

If we set $\alpha = \epsilon$ and $\epsilon' = \epsilon(1+\epsilon)$, we achieve an $(1+\epsilon')$-approximation.
To analyze the running time, we only need to count the number of vertices of the refined set of grids.
There are at most $O(\min(n, \beta))$ grids created, each has $O(M^2) = O\Big(\frac{n^2}{\epsilon'^2}\Big)$ grid points.
In total, we have $O\Big(\min (n, \beta) \frac{n^2}{\epsilon'^2} \Big)$ vertices.
Applying the algorithm from \cite{carvalho2021fast} for the graph version (with a running time of $O(n |V|\log |V|+ n|E|)$ where $n$ is the number of drones, $|V|$ and $|E|$ are the numbers of the graph's vertices and edges), we can compute our solution in time $O\Big(\min (n,  \beta)^2  \frac{n^5}{\epsilon'^4} \Big)$.
This is because in our case, we have a complete graph, so the term $O(n|E|) = O(n|V|^2)$ dominates the running time.

\begin{theorem}
    There exists an FPTAS for \textsc{FD} in the plane.
\end{theorem}

We can now tweak the above FPTAS so that it works with the presence of polygonal obstacles in the plane.
First, we need to bound the movements of the drones differently since we cannot reuse $\tau^{(2)}$.
Consider the solution where only a single drone is used.
The optimal solution with this restriction is a 3-approximation (See Appendix~\ref{app:3approx}).
Let $\tau^{(3)}$ be the optimal makespan for the single-drone restriction.
For each drone, we consider the geodesic disk at $s$ with radius $\tau^{(3)}v_i$ (a geodesic disk is defined similarly to a normal disk in the plane, but with geodesic distance between two points: $d(a,b)$ is the length of the shortest path from $a$ to $b$ in the domain with obstacles).

\begin{figure}
    \centering
    \includegraphics[width=0.85\linewidth]{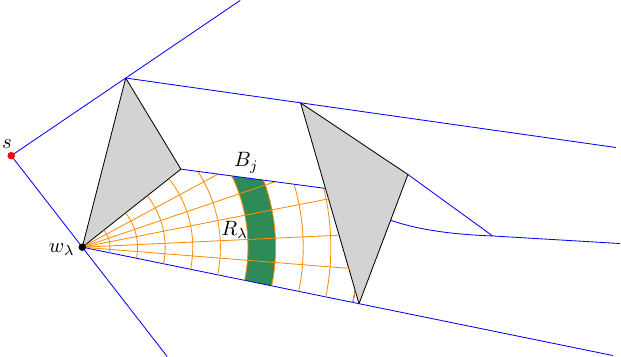}
    \caption{The shortest path map in the domain where each cell is further divided into smaller cells by rays and bands from $w_{\lambda}$.
    Obstacle edges may truncate grid cells.
    The boundary of each region $R_\lambda$ (blue) may also truncate grid cells, but this boundary is not a physical obstacle: drones can freely cross it. Therefore, snapping an exchange point to a grid point near a region boundary incurs the same error as snapping within the interior of a region, and the cell diameter bound applies unchanged.}
    \label{fig:obstacles}
\end{figure}
Next, we compute the shortest path map \cite{hershberger1999optimal} within this geodesic disk; this partitions the disk into $O(N)$ combinatorial regions $R_\lambda$, where each point in $R_\lambda$ has the same sequence of vertices on the shortest path from $s$.
For each region $R_\lambda$, denote $w_\lambda$ as the last vertex in the shortest path from $s$.
We split each region $R_\lambda$ angularly and radially around $w_\lambda$ as follows.
For the angular split, we shoot a ray from $w_\lambda$ through every obstacle vertex visible to $w_\lambda$ within $R_\lambda$, as well as rays along the obstacle edges incident to $w_\lambda$.
This partitions $R_\lambda$ into convex pieces; across all regions, this creates $O(N)$ pieces in total since the number of such visibility rays is proportional to the complexity of the shortest path map.
Within each piece, we further partition it into $M$ smaller subpieces by shooting $M-1$ rays from $w_\lambda$ with uniform angular spacing.
For the radial split, we partition each subpiece into $M$ radial bands $B_j=\big\{p \mid \frac{j-1}{M}\tau^{(3)}v_i \le d(p, w_\lambda)  \le \frac{j}{M}\tau^{(3)}v_i\big\}$ for $1\le j \le M$.
The boundary of each such band consists of arc segments together with some subsets of obstacle edges.
Both splitting types give $O(NM^2)$ cells per drone.

\begin{lemma}
    Each cell has a geodesic diameter of at most $\frac{\tau^{(3)}v_i(1+2\pi)}{M}$.
\end{lemma}

\begin{proof}
    With the construction above, each cell is a simple closed curve whose boundary is composed of $O(1)$ components: at most two ray segments from $w_{\lambda}$, at most two arcs from two concentric circles, and at most one connected subset of an edge of an obstacle.
    The upper bound of one for the edge subset component holds because each convex piece is bounded by two consecutive visibility rays from $w_{\lambda}$; such a piece is either a cone (with apex $w_{\lambda}$) or a triangle when there is exactly one obstacle edge spanning the cone (*).
    Therefore, cells acquired by further splitting (angularly and radially) within each piece will interact with at most one edge of some obstacle.
    Note that rays collinear with obstacle edges are considered as rays, not obstacle edges; this does not change our analysis.

    Consider a cell with no edge subset component: the cell has a radial length of $\frac{\tau^{(3)}v_i}{M}$ and outer arc length at most $\frac{2\pi\tau^{(3)} v_i}{M}$.
    We call such a cell a free cell.
    Its geodesic diameter is at most $\frac{\tau^{(3)}v_i(1+2\pi)}{M}$: any two points in a free cell can be connected by a path along the arcs and rays within the cell, which is obstacle-free.
    For a non-free cell, there is at most one subset of an obstacle edge $e$ on the boundary.
    A non-free cell results from intersecting a free cell with the half-plane tangent to $e$ containing $w_{\lambda}$.
    Since a free cell is a simple closed curve with one concave arc, the line containing $e$ intersects the free cell in at most 2 connected components, each of which we treat as a separate cell.
    Each such component is connected and obstacle-free, so any two points in it can be joined by a path within the component of length at most $\frac{\tau^{(3)}v_i(1+2\pi)}{M}$.
    Therefore, any non-free cell also has a geodesic diameter bounded by $\frac{\tau^{(3)}v_i(1+2\pi)}{M}$.
\end{proof}
We note that even when an edge can split a free cell into two, the total number of cells is still $O(NM^2)$: because of (*), each free cell can only be split by an edge at most once. Using the drone-pruning process similar to the FPTAS for obstacle-free domains, one grid per drone, we obtain a solution $S' = (\Pi', Q')$.

With the circular grid, snapping both exchange points ($a_j$ and $b_j$) of a drone $\pi'_j$ to two grid vertices will add a time of at most $\frac{\tau^{(3)}2(1+2\pi)}{M}$.
Then, the total error is bounded by
\[
\sum_{j=k}^{\mu} \frac{\tau^{(3)}2(1+2\pi)}{M} \le \frac{\tau_{S^*}(m)m6(1+2\pi)}{M}.
\]
We can get an $(1+\epsilon)$-approximation by setting $M \ge \frac{6n}{\epsilon} (1+2\pi)$.
With one grid for each drone, this gives us $O\Big(\frac{n^3N}{\epsilon^2}\Big)$ grid vertices in total.
\begin{corollary}
    There exists an FPTAS for \textsc{FD} in the plane with obstacles.
\end{corollary}

\subsection{Algebraic hardness for \texorpdfstring{$n=2$}{n=2}}
\begin{figure}
    \centering
    \includegraphics[width=0.65\textwidth]{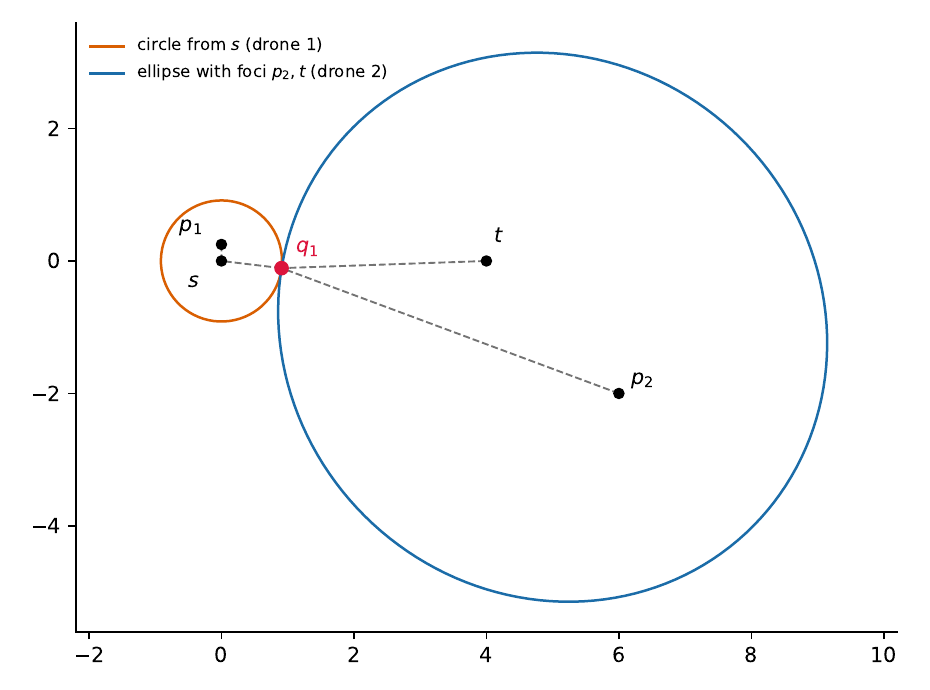}
    \caption{The instance that shows algebraic hardness. The red dot shows the (approximated) location where the handoff occurs between drone 1 and drone 2 in the optimal solution. The true location of this exchange point is the root of an unsolvable polynomial.}
    \label{fig:alg_hardness}
\end{figure}
In this section, we prove that \textsc{FD} is algebraically hard.
A problem is considered algebraically hard when there exists an instance of the problem where the inputs are only rationals and solving it optimally is equivalent to finding the root(s) of some unsolvable polynomials, i.e., we cannot express these roots with radicals ($\sqrt[x]{\cdot}$).
We will demonstrate that such an instance exists.
Consider an instance with two drones: drone 1 with speed $v_1 = 1/2$ at
$p_1 = (0, 1/4)$, and drone 2 with speed $v_2 = 7/3$ at $p_2 = (6, -2)$.
The package is at $s = (0,0)$ and the target is at $t = (4,0)$.
Fig.~\ref{fig:alg_hardness} illustrates this instance.
To solve this, we formulate the following optimization problem:
\begin{align*}
\min_{q_1} \;\;\; & \frac{1}{v_2}(d(p_2,q_1) + d(q_1,t)) \\
\textnormal{s.t.} \;\;\; & \frac{1}{v_1}(d(p_1,s)+d(s,q_1))=\frac{1}{v_2}d(p_2,q_1).
\end{align*}
Note that this assumes that OPT uses both drones (which we can verify numerically by finding an arbitrary two-drone solution that is strictly better than both single-drone solutions).
This optimization problem is equivalent to growing an ellipse from the two foci $p_2$ and $t$ with speed $v_2$ and growing a circle (with a delayed start to account for $\frac{1}{v_1}d(p_1,s)$) from the center $s$ with speed $v_1$ until both curves touch (at $q_1$).
The tangent point of the two curves will be the optimal exchange point.
Attempting to solve the above gives us the following polynomial whose roots include the optimal $x$-coordinate of $q_1$:
\begin{align*}
f(x) = 18879804846596656133000505 \\
+ 150710862885638628505994880x-901032721281600508602112592x^2\\
-4056928330895527441141838336x^3 +14595856936709624502012609600x^4 \\
-2378132853295078134017314816x^5-13600158548889330219566628864x^6\\
+2471994260131460539419197440x^7+3318801307437382496577126400x^8.
\end{align*}
For our proof, we will employ the framework introduced in Lemma~8 in \cite{bajaj1988algebraic}.
A necessary condition for this framework is that the degree of the polynomial must be at least $6$ and even, which the above satisfies.
Then, if the degree of a polynomial is even, we have to show that there exist 3 ``good'' primes such that (1) they do not divide the discriminant of the polynomial and (2) a $\deg (f)$-cycle, a $(\deg (f)-1)$-cycle, and a permutation of the type $2+(\deg (f)-3)$ occur when we factorize the polynomial modulo these primes (here, $\deg (f)$ is the degree of $f(x)$).
The three primes we found that work are 277, 853, and 1049, giving us the following:
\begin{align*}
\begin{split}
f(x) &\equiv 85 (x^8 + 122x^7 + 104x^6 + 264x^5 + 148x^4 + 244x^3 + 35x^2 + 166x + 212)  \\ &\pmod {277} \\
f(x) &\equiv 560 (x + 739)(x^7 + 445x^6 + 821x^5 + 716x^4 + 732x^3 + 664x^2 + 22x + 656)  \\ &\pmod {853} \\
f(x) &\equiv 499 (x + 546)(x^2 + 517x + 680)(x^5 + 227x^4 + 158x^3 + 600x^2 + 423x + 639)  \\ &\pmod {1049}.
\end{split}
\end{align*}

This shows existence of an $\deg (f)$-cycle, an $(\deg (f)-1)$-cycle, and a permutation
of the type $2+(\deg (f)-3)$; by \cite{bajaj1988algebraic}, $\mathrm{Gal}(f)$
is not solvable.
Note that $277$ does not divide the leading coefficient and that $f(x) \bmod 277$ is irreducible of degree $8$, so $f(x)$ admits no lower-degree factorization over $\mathbb{Z}$; since $f(x)$ is primitive, it is irreducible over $\mathbb{Q}$.
All roots of $f(x)$ therefore share a minimal polynomial with a non-solvable Galois
group, and none of them (in particular the $x$-coordinate of the tangency point) is expressible in radicals.
The Python script for this instance is shown in Appendix~\ref{app:mathematica}.

\section{Energy-efficient delivery}\label{sect:ED}

We provide an FPTAS for energy-efficient delivery (\textsc{ED}): here, each drone has a different energy consumption rate $r_i$ ($r_1 < r_2 < \ldots < r_n$; here, drone $1$ is the most efficient drone), and we want to minimize the total amount of energy used in the schedule.
The total energy spent in an optimal solution $S^*$ can be written as:
\[
\phi_{S^*}(m) = \sum_{j=1}^m r_{\pi_j}(d(p_{\pi_j}, q_{j-1})  +d(q_{j-1}, q_{j})).
\]

We will use a grid discretization strategy similar to what was used previously for \textsc{FD}. 
The set of grids is created based on a set of intervals for energy consumption rates $I_r=\{[r_1(1+\alpha)^\ell, r_1(1+\alpha)^{\ell+1}]| 0 \le \ell < \beta \}$ where $\beta = \big \lceil \log_{1+\alpha}\frac{r_n}{r_1} \big \rceil $.
To decide on the bound of the grid, we again need an upper bound.
Consider the problem where only one drone is allowed to carry the package (no handoff), and let $\phi^{(2)}$ be the optimal value of this version; then $\phi^{(2)} \le 2\phi_{S^*}(m)$ (see Appendix~\ref{app:phi_1} for proof).

For each bound $r_1(1+\alpha)^{\ell}$ of a non-empty interval in $I_r$, we create a grid centered at $s$ in a box of side length $L_{\ell}=\frac{2\phi^{(2)} }{r_1(1+\alpha)^{\ell}}$ and $M \times M$.
We claim that the optimal exchange points of drone $\pi_j$ belong to the bounding box of the grid with length $L_{\ell}$. 
First, see that the package moves toward lower energy consumption rates, i.e., $\forall i< j:r_{\pi_i} > r_{\pi_j}$.
A hypothetical route with $\pi_j$ moving the package from $s$ to $q_{j-1}$ has energy
\[r_{\pi_j}d(s, q_{j-1}) \le r_{\pi_j}\sum_{i<j} d(q_{i-1}, q_i) \le \sum_{i<j} r_{\pi_i}d(q_{i-1}, q_i) \le \phi_{S^*}(m) \le \phi^{(2)}.\]
Which means $d(s, q_{j-1}) \le \frac{\phi^{(2)}}{r_{\pi_j}} \le \frac{\phi^{(2)}}{r_1(1+\alpha)^{\ell}} = \frac{L_{\ell}}{2}$.
Applying a similar argument:
\[d(s, q_j) \le d(s, q_{j-1}) + d(q_{j-1}, q_j) \le \frac{1}{r_{\pi_j}}\sum_{i\le j} r_{\pi_i} d(q_{i-1}, q_i) \le \frac{\phi^{(2)}}{r_{\pi_j}} \le \frac{L_{\ell}}{2}.\]
To prove the correctness of using the grid set, first see that the additional energy caused by snapping an exchange point of drone $\pi_j$ to a grid vertex is at most $\frac{L_{\ell} r_{\pi_j} \sqrt{2}}{M}  \le \frac{\phi^{(2)}(1+\alpha) 2\sqrt{2}}{M}$.
When we snap the two optimal exchange points of drone $\pi_j$ to the grid, the act of moving $q_{j-1}$ perturbs both $d(p_{\pi_j}, q_{j-1})$ and $d(q_{j-1}, q_j)$, while moving $q_j$ perturbs only the latter. 
Therefore, by triangle inequality, the additional energy cost for each drone is at most $\frac{\phi^{(2)}(1+\alpha) 6\sqrt{2}}{M}$.
The total energy after snapping is now:
\[
\phi_{S^*}(m) + \sum_{j=1}^m \frac{\phi^{(2)}(1+\alpha) 6\sqrt{2}}{M} \le \phi_{S^*}(m)\Bigg(1+ \frac{m(1+\alpha)12\sqrt{2}}{M}\Bigg).
\]
We get an FPTAS by setting $M = \Big\lceil \frac{n12\sqrt{2}}{\epsilon} \Big\rceil, \alpha = \epsilon$ and $\epsilon' = \epsilon(1+\epsilon)$.
Using the $O(n + |V|^3)$ algorithm for \textsc{ED} in a graph \cite{bartschi2017energy}, the FPTAS running time is $O\Big(\min(n,\beta)^3 \frac{n^6}{\epsilon'^6}\Big)$.
\begin{theorem}
    There exists an FPTAS for \textsc{ED} in the plane.
\end{theorem}

\section{Conclusion and open problems}
We presented various results on two variants of the package delivery problem with collaborative drones in the plane.
A primary focus was on establishing the computational complexity of \textsc{FD}.
On the algorithmic results, a constant-factor approximation and an FPTAS were provided for \textsc{FD}.
The FPTAS can be modified to work in domains with obstacles, and with an energy-efficient objective.
For future work, we want to investigate whether \textsc{FD} or \textsc{ED} is NP-hard.
We also want to study the energy-bounded version of the problem where each drone has an energy limit, and they can not travel more than allowed: in this case, deciding if the package is deliverable is also an important question.


\bibliographystyle{plainurl}
\bibliography{ref}

\newpage
\appendix

\input{appendix}

\end{document}

%% file: related_work.tex
The delivery problem with collaborative drones on graphs was first studied in \cite{chalopin2013data} by Chalopin et al.
In this version, drones can have different energy consumption rates, and the goal was to minimize total energy burned.
In a later work \cite{bartschi2018collective}, among other variants, the authors presented fast delivery: a slightly different formulation allows drones to have varied speeds instead, and the task is to compute a schedule with minimum time.
Both versions are known to be polynomial with a running time of $O(n +|V|^3)$ (for energy minimization) and $O(n^2|E| + n|V|^2 + APSP)$ (for time minimization).
Here, $n$ is the number of drones, $|V|$ and $|E|$ are the number of vertices and edges of the input graph, and $APSP$ is the running time of all-pair shortest paths.
The running time for fast delivery was later reduced to $O(n|V|\log|V| + n|E|)$ by Carvalho et al.  \cite{carvalho2021fast}.
For fast delivery in the plane, \cite{coleman2021message} provided a $(\sqrt{2} + \epsilon)$-approximation with a running time of $O\Big(n^3\Big(\frac{\Delta}{\epsilon} \Big)^2 \log\Big(n\frac{\Delta}{\epsilon} \Big)\Big)$, where $\Delta$ is a positive parameter depending on the diameter of the input locations.

With more complicated objectives or additional constraints, the problem often becomes NP-hard.
For instance, in \cite{bartschi2018collective}, the authors proved that both lexicographically minimizing an objective tuple of (time, energy), as well as optimizing a weighted sum of time and energy consumption, are NP-hard.
They also showed that the latter objective can be approximated within a factor of 3 by using one drone.
One realistic constraint is putting the budget on the energy or the distance each drone can expend during travel.
Some works showed that the problem is NP-hard in graphs \cite{bartschi2020collaborative}, even if all drones have the same budget  \cite{chalopin2013data}.
Even more surprisingly, Chalopin et al. \cite{chalopin2014data} proved that the problem is weakly NP-hard when all drones, and $s$ and $t$, are on the $x$-axis, and they give a quasi-pseudopolynomial time algorithm.
If we restrict the regions where drones can operate and optimize for time, then the problem also becomes NP-hard even on path graphs \cite{bartlmae2026parameterized}.
In more general graphs, Erlebach et al. also demonstrated that it is hard to approximate within a ratio of $O(n^{1-\epsilon})$ \cite{erlebach2022package}.
In the case where the domain is the 2D plane, but package exchanges are limited to a straight line segment (not necessarily the $s-t$ line segment), then the problem can be solved in polynomial time \cite{blank2025algorithms}.

When the input has multiple packages and targets, it is easy to see that the problem is NP-hard from the traveling salesman problem.
Carvalho et al. proved that finding the schedule with minimum makespan is NP-hard with just two packages \cite{carvalho2021fast}.
If the package sources and targets are well-separated and no collaboration is allowed, then the problem is the same as the classic stacker crane problem \cite{frederickson1978approximation}, which has an $\frac{9}{5}$-approximation algorithm for a single-drone team, and $\frac{14}{5}- \frac{1}{n}$ if $n$ same-speed drones are involved.

%% file: appendix.tex
\section{Omitted proofs of Section~\ref{sect:FD}}

\subsection{Proof of Lemma~\ref{lem:canonical}}\label{app:canonical}

\canonical*

Since (i) is always possible, we will only prove existence for (ii).
    Let $S$ be an optimal solution and let $j$ be the smallest index at which (ii) fails; we construct an optimal $S'$ agreeing with $S$ on all handoffs before $j$, satisfying (ii) at $j$, and with $\tau_{S'}(\ell) \le \tau_S(\ell)$ for all $\ell \ge j$. Repeating the same argument over $j = 2, \dots, m$ will yield a canonical optimal solution.

    Suppose first that $\pi_{j-1}$ reaches $q_{j-1}$ at time $\tau$ and waits for $\pi_j$, which arrives at $\tau' > \tau$. Since $\pi_j$ travels along the segment $p_{\pi_j}q_{j-1}$, letting $\pi_{j-1}$ continue toward $\pi_j$ produces a meeting point $q'_{j-1}$ on that segment. 
    As $q'_{j-1}$ lies on $p_{\pi_j}q_{j-1}$, the triangle inequality gives
    \[
        d(p_{\pi_j}, q'_{j-1}) + d(q'_{j-1}, q_j)
        \;\le\; d(p_{\pi_j}, q_{j-1}) + d(q_{j-1}, q_j),
    \]
    so $\pi_j$ reaches $q_j$ no later than in $S$.

    Otherwise $\pi_j$ reaches $q_{j-1}$ at time $\tau'$ and waits for the carrier $\pi_{j-1}$, which arrives at $\tau > \tau'$. Let $\pi_j$ instead fly back along $\pi_{j-1}$'s trajectory and meet it at $q'_{j-1}$ at time $\tilde{\tau} < \tau$, where $d(q_{j-1}, q'_{j-1}) = v_{\pi_{j-1}}(\tau - \tilde{\tau})$. Then $\pi_j$ reaches $q_j$ at time
    \[
        \tilde{\tau} + \tfrac{1}{v_{\pi_j}} d(q'_{j-1}, q_j)
        \;\le\; \tilde{\tau}
        + \tfrac{v_{\pi_{j-1}}}{v_{\pi_j}}(\tau - \tilde{\tau})
        + \tfrac{1}{v_{\pi_j}} d(q_{j-1}, q_j)
        \;\le\; \tau + \tfrac{1}{v_{\pi_j}} d(q_{j-1}, q_j),
    \]
    using $v_{\pi_{j-1}} \le v_{\pi_j}$, so again $\pi_j$ reaches $q_j$ no
    later than in $S$.

    In both cases the trajectories of $\pi_1, \dots, \pi_{j-1}$ before $q_{j-2}$ are unchanged, so (ii) still holds at every index below $j$,
    and every subsequent drone's constraint is unchanged or relaxed. In particular, the makespan does not increase, so $S'$ is also optimal.

\subsection{The package does not move toward the negative direction of the $x$-axis in $S^{(2)}$}\label{app:claim1}

\neverbackward*

\begin{proof}
To prove the claim, we first define forward and backward drones.
When a drone takes the package on the $x$-axis and moves it toward $t$, we call it a forward drone.
Otherwise, we call it a backward drone.
Let there be some solution $S$ in which there is at least one backward drone.
Notice that whenever a forward drone $\pi_j$ hands the package to a backward drone $\pi_{j+1}$ at $q_j$, there must be another forward drone $\pi_\ell: \ell > j+1$ that crosses over $q_j$ later: otherwise, the package could not be delivered to $t$.
Because of this, we can simply remove all drones from $\pi_{j+1}$ to $\pi_{\ell-1}$, let $\pi_j$ wait at $q_j$ until $\pi_\ell$ arrives, and the makespan stays the same.
We can apply the same process repeatedly until we have a forward drone-only solution.
\end{proof}

\subsection{Optimal solution with single-drone restriction is a 3-approximation}\label{app:3approx}

In $S^*$, the makespan is equal to the travel time of the last drone $\pi_m$.
If we force $\pi_m$ to go to $s$ $\big(\frac{1}{v_{\pi_m}}d(p_{\pi_m},s)\big)$ and then to $t$ $\big(\frac{1}{v_{\pi_m}}d(s,t)\big)$, due to the triangle inequality and Lemma~\ref{lem:never-too-far} (which is still true in the presence of obstacles):
\begin{align*}
    \frac{1}{v_{\pi_m}}d(p_{\pi_m},s) + \frac{1}{v_{\pi_m}}d(s,t) \le &\frac{1}{v_{\pi_m}}d(p_{\pi_m},q_{m-1}) + \frac{1}{v_{\pi_m}}d(q_{m-1}, s) +\frac{1}{v_{\pi_m}}d(s,t) \\ \le & 3\tau_{S^*}(m)
\end{align*}
Note that $\frac{1}{v_{\pi_m}}d(s,t) \le \tau_{S^*}(m)$ since the package trajectory in any delivery has a length of at least $d(s,t)$, and no drone in $S^*$ has speed faster than $v_{\pi_m}$.

\section{Supplementary materials for algebraic hardness}\label{app:mathematica}

We use Python with sympy to identify the required polynomials. 
First, we define the delivery instance and the optimization problem:
\lstset{
    frame=tb, 
    tabsize=4, 
    showstringspaces=false, 
    numbers=left, 
    commentstyle=\color{gray}, 
    keywordstyle=\color{blue}, 
    stringstyle=\color{ForestGreen}, 
    xleftmargin=2em,frame=single,framexleftmargin=0.0em
}
\begin{lstlisting}[language=Python,breaklines]
import sympy as sp
import mpmath as mp

mp.mp.dps = 50

x, y, D0, D1, D2, lam = sp.symbols('x y D0 D1 D2 lam')

# --- instance ----
S  = (sp.Integer(0), sp.Integer(0))
T  = (sp.Integer(4), sp.Integer(0))
P1 = (sp.Integer(0), sp.Rational(1, 4));  V1 = sp.Rational(1, 2)
P2 = (sp.Integer(6), sp.Integer(-2));     V2 = sp.Rational(7, 3)

dist = lambda a, b: sp.sqrt((a[0] - b[0])**2 + (a[1] - b[1])**2)

C0 = sp.nsimplify(dist(P1, S))            # head-start distance d(p1, s)
assert C0.is_rational and C0 != 0, "d(p1,s) must be rational and nonzero"

n0 = sp.expand((x - S[0])**2 + (y - S[1])**2)     # d(q,s)^2
n1 = sp.expand((x - P2[0])**2 + (y - P2[1])**2)   # d(q,p2)^2
nt = sp.expand((x - T[0])**2 + (y - T[1])**2)     # d(q,t)^2
RELS = [(D0**2 - n0, D0), (D1**2 - n1, D1), (D2**2 - nt, D2)]
\end{lstlisting}

The below script find the exact representation of $f(x)$:

\begin{lstlisting}[language=Python,breaklines]
def reduce_radicals(expr):
"""Reduce to degree <= 1 in each Di using Di^2 = (polynomial in x, y)."""
for rel, D in RELS:
    sub = sp.solve(rel, D**2)[0]
    p = sp.Poly(expr, D)
    expr = sp.expand(sum(c * sub**(k // 2) * D**(k % 2)
                         for (k,), c in zip(p.monoms(), p.coeffs())))
return expr

# Objective  (D1 + D2)/v2   subject to   (d(p1,s) + D0)/v1 = D1/v2.
# Lagrange conditions, multiplied through by v1*v2*D0*D1*D2:
Lx = (V1*D0*((x - P2[0])*D2 + (x - T[0])*D1)
      - lam*(V2*(x - S[0])*D1*D2 - V1*(x - P2[0])*D0*D2))
Ly = (V1*D0*((y - P2[1])*D2 + (y - T[1])*D1)
      - lam*(V2*(y - S[1])*D1*D2 - V1*(y - P2[1])*D0*D2))

E = sp.expand(sp.resultant(sp.Poly(Lx, lam), sp.Poly(Ly, lam)))   # eliminate lambda
for _ in range(4):
    E = reduce_radicals(E)

# The constraint makes D1 affine in D0; with D0^2 = n0 and D1^2 = n1 this
# forces D0 (hence D1) to be a polynomial in x, y, only D2 stays a radical.
K = V2 / V1
D0p = sp.expand(sp.simplify((n1 - K**2*C0**2 - K**2*n0) / (2*C0*K**2)))
D1p = sp.expand(K * (C0 + D0p))
CURVE = sp.expand((D0p**2 - n0) * sp.denom(sp.together(D0p**2 - n0)))

Es = sp.expand(sp.numer(sp.together(E.subs({D0: D0p, D1: D1p}))))
A, B = sp.expand(Es.coeff(D2, 0)), sp.expand(Es.coeff(D2, 1))
assert sp.expand(Es - A - B*D2) == 0
ESQ = sp.expand(A**2 - B**2 * nt)      # <-- squaring: extraneous roots enter here

RES = sp.Poly(sp.resultant(sp.Poly(ESQ, x, y), sp.Poly(CURVE, x, y), y), x)
f = [g for g, _ in sp.factor_list(RES.as_expr())[1]
     if sp.Poly(g, x).degree() == 8][0]
P = sp.Poly(f, x)

print("=" * 76)
print("(A)  f(x), derived exactly")
print("=" * 76)
print(f"degree {P.degree()} | content {P.content()} | irreducible over Q: {P.is_irreducible}")
for k, c in enumerate(P.all_coeffs()[::-1]):
    print(f"  x^{k}: {c}")
\end{lstlisting}

To prove that the optimal exchange point lies on the tangent of the two curves, we need to conclude that the single-drone solutions (drone 1 or drone 2 picks up the package from $s$ and goes directly to $t$) are strictly worse than a solution with an exchange between drone 1 and 2.
The script below show this (numerically):
\begin{lstlisting}[language=Python,breaklines]
d = lambda a, b: sp.sqrt((a[0]-b[0])**2 + (a[1]-b[1])**2)
s, t, p1, p2, v1, v2 = (0,0), (4,0), (0,sp.Rational(1,4)), (6,-2), sp.Rational(1,2), sp.Rational(7,3)
mk = lambda q: sp.Max((d(p1,s)+d(s,q))/v1, d(p2,q)/v2) + d(q,t)/v2   # makespan with handoff at q
alone1, alone2, wit = (d(p1,s)+d(s,t))/v1, mk(s), mk((sp.Rational(9,10), sp.Rational(-1,10)))
for n, e in (("drone 1 alone", alone1), ("drone 2 alone", alone2), ("witness q=(9/10,-1/10)", wit)):
    print(f"{n:24} = {sp.nsimplify(e)} = {float(e):.10f}")
print("witness < drone 1 alone:", sp.ask(sp.Q.positive(sp.simplify(alone1 - wit))))
print("witness < drone 2 alone:", sp.ask(sp.Q.positive(sp.simplify(alone2 - wit))))
\end{lstlisting}

Finally, we find the 3 representations of $f(x)$ after taking its modulo with the good primes that were found manually:
\begin{lstlisting}[language=Python,breaklines]
lead, disc = P.all_coeffs()[0], sp.discriminant(P)
print()
print("=" * 76)
print("(C)  Irreducibility over Q and the good-prime conditions")
print("=" * 76)
print(f" content(f) = {P.content()}  (primitive)")
print(f" f irreducible over Q: {P.is_irreducible}")
for p in (277, 853, 1049):
    degs = sorted(int(sp.Poly(h, x).degree())
                  for h, m in sp.factor_list(f, modulus=p)[1] for _ in range(m))
    print(f"  p = {p:5d} | p does not divide lead: {lead % p != 0!s:5} "
          f"| p does not divide disc: {disc % p != 0!s:5} | factor degrees: {degs}")
    print(sp.factor_list(f, modulus=p, symmetric=False))
\end{lstlisting}

\section{Proof that using a single drone optimally gives a 2-approximation for ED}\label{app:phi_1}
Consider the optimal solution $S^* =  (\Pi^*, Q^*)$, its total energy cost is:
\[
\phi_{S^*}(m) =  \sum_{j=1}^m r_{\pi_j}(d(p_{\pi_j}, q_{j-1}) + d(q_{j-1}, q_j)) = \sum_{j=1}^m r_{\pi_j}d(p_{\pi_j}, q_{j-1}) + \sum_{j=1}^m r_{\pi_j}d(q_{j-1}, q_j)
\]
We know that the energy consumption rate $r_j, j <m,$ of all drones that carry the package from $s$ to $q_{m-1}$ is worse than $\pi_m$, therefore:
\[
 r_m\sum_{j=1}^{m-1} d(q_{j-1},q_j) \le \sum_{j=1}^{m-1} r_j d(q_{j-1},q_j)
\]
Adding $r_md(p_m, q_{m-1})$ on both sides:
\[
 r_m\Bigg(d(p_m, q_{m-1}) + \sum_{j=1}^{m-1} d(q_{j-1},q_j) \Bigg)\le r_md(p_m, q_{m-1})+\sum_{j=1}^{m-1} r_j d(q_{j-1},q_j) \le \phi_{S^*}(m) 
\]
Furthermore, due to the triangle inequality, we have
\[
d(p_m, s) \le d(p_m, q_{m-1}) + \sum_{j=1}^{m-1} d(q_{j-1},q_j),
\]
and
\[
d(s,t) \le \sum_{j=1}^{m} d(q_{j-1},q_j)
\]
If we use drone $m$ to pick up the package at $s$ and deliver it to $t$, then the total energy cost is:
\begin{align*}
r_m(d(p_m, s) + d(s,t)) &\le r_m\Bigg(d(p_m, q_{m-1}) + \sum_{j=1}^{m-1} d(q_{j-1},q_{j}) +  \sum_{j=1}^{m} d(q_{j-1},q_j)\Bigg)\\
&\le 2\phi_{S^*}(m) 
\end{align*}

We note that using this argument for \textsc{FD} will give a 3-approximation instead.